# YouTube QoE Evaluation Tool for Android Wireless Terminals

**Gerardo Gómez**[1], **Lorenzo Hortigüela**[2], **Quiliano Pérez**[2], **Javier Lorca**[2], **Raquel García**[2], **Mari Carmen Aguayo-Torres**[1]

[1]*Department of Communications Engineering, University of Malaga, 29071 Malaga, Spain*

[2]*Telefonica I+D, 28006 Madrid, Spain*

E-mail addresses: ggomez@ic.uma.es, lhm@tid.es, qpt@tid.es, jlh@tid.es, rgp@tid.es, aguayo@ic.uma.es

Correspondence should be addressed to Gerardo Gómez, ggomez@ic.uma.es

**Abstract** – In this paper, we present an Android application which is able to evaluate and analyze the perceived Quality of Experience (QoE) for YouTube service in wireless terminals. To achieve this goal, the application carries out measurements of objective Quality of Service (QoS) parameters, which are then mapped onto subjective QoE (in terms of Mean Opinion Score, MOS) by means of a utility function. Our application also informs the user about potential causes that lead to a low MOS as well as provides some hints to improve it. After each YouTube session, the users may optionally qualify the session through an online opinion survey. This information has been used in a pilot experience to correlate the theoretical QoE model with real user feedback. Results from such an experience have shown that the theoretical model (taken from the literature) provides slightly more pessimistic results compared to user feedback. Users seem to be more indulgent with wireless connections, increasing the MOS from the opinion survey in about 20% compared to the theoretical model, which was obtained from wired scenarios.

**1.    Introduction**

Real-time entertainment services (comprised mostly of streaming video and audio) are becoming one of the dominant web-based services in telecommunications networks. In particular, YouTube service is currently the largest single source of real-time entertainment traffic and the third most visited Internet site (preceded by Google and Facebook). It has emerged to account for more Internet traffic than any other service. Mobile networks have the highest proportion of real-time entertainment traffic. Nowadays, YouTube leads the way, accounting for 20-25% of total traffic in mobile networks. Additionally, 27.8% of all YouTube traffic (first half 2012) has been consumed on a Smartphone or tablet [1].

The combination of increasing device capabilities, high-resolution content and longer video duration (largely due to live content) means that YouTube's growth will continue for the foreseeable future. Driven by higher bitrates and enhanced capabilities of mobile devices, the trend is also going towards High Definition (HD) video, which considerably enhances quality demand. That is the reason mobile networks operators are following this trend, as it will be hugely influential on network requirements and subscriber Quality of Experience (QoE).

The QoE has been usually evaluated through subjective tests carried out on the users in order to assess their degree of satisfaction with Mean Opinion Score (MOS) indicator [3]. This type of approach is obviously quite expensive, as well as annoying to the user. That is why in recent years new methods have been used to estimate the QoE based on certain performance indicators associated with services. The evaluation methodology used by most network operators to obtain statistical QoE is based on field testing. These tests often use mobile handsets as a modem, with laptop computers that perform the tests and keep statistics. However, this process is expensive in terms of resources and staff, and also it does not use the entire protocol stack implemented in the terminal. These drawbacks are solved by integrating QoE analyzers in the mobile terminal itself so that measurements of statistics are specific to each terminal. Thus, additional measurements can be collected (along the protocol stack) to allow for enhanced analysis of the performance of each service. Furthermore, if mobile terminals are able to report the measurements to a central server, the QoE assessment process is simplified significantly.

Recently, a number of works have focused on developing subjective QoE evaluation frameworks for mobile users. For instance, an implementation of a QoE measurement framework on the Android platform is presented in [4][5], although results are limited to a laboratory environment. The works in [6][7] present a framework for measuring the QoE for distorted videos in terms of Peak Signal-to-Noise Ratio (PSNR) or a modified metric called cPSNR, respectively. A QoE framework for multimedia services (named as QoM) for run time quality evaluation of video streaming services is presented in [8]; this approach is based on the influence of QoE factors, various network and application level QoS parameters, although no evaluation of the proposed framework in a context of real wireless network has been performed. In [11], the problem of YouTube QoE monitoring from an access provider's perspective is investigated, showing that it is possible to detect application-level stalling events by using network-level passive probing only. The work in [12] describes a tool which constantly monitors the YouTube application comfort, making it possible to estimate the time when the YouTube player is stalling.

Other works are focused on specific YouTube models to compute the QoE. In [9][10], different QoE YouTube models that take into account the key influence factors (such as stalling events caused by network bottlenecks) in the quality perception are presented. They quantify the impact of initial delays on the user perceived QoE by means of subjective laboratory and crowdsourcing studies. Other works are devoted to estimate the

MOS for video services [14][4][22]; among them, the analysis presented in [22] provides a utility function for HTTP video streaming as a function of three application performance metrics: initial buffering time, mean rebuffering time and re-buffering frequency.

However, none of previous works have performed a deep validation of existing models through real tests over different radio technologies. In this work we describe an Android application that carries out measurements of objective Quality of Service (QoS) indicators associated to YouTube service; this performance indicators are then mapped onto subjective QoE (in terms of MOS). Our application also informs the user about possible causes that lead to a low MOS as well as provides some hints to improve it. After each YouTube session, the users may optionally qualify the session through an opinion survey. This information has been used in a pilot experience to correlate the theoretical QoE model with real user feedback.

The remainder of this paper is structured as follows. A description of the YouTube QoE evaluation method is given in section 2, specifying its main performance indicators. In section 3, we describe our Android application for YouTube QoE evaluation. The results from a YouTube evaluation pilot experience are analyzed in section 4. Finally, some concluding remarks are given in section 5.

## 2. YouTube QoE evaluation method

YouTube service employs progressive download technique, which enables the playback of the video before the content downloaded is completely finished [13]. Old YouTube delivery service for mobile terminals (through the mobile YouTube link http://m.youtube.com) was based on conventional video streaming architecture, i.e. Real Time Streaming Protocol (RTSP) and Real Time Protocol (RTP), being the latter transported over User Datagram Protocol (UDP). However, current delivery service (both for Smartphone and PCs) uses progressive video download via HyperText Transfer Protocol (HTTP) over Transmission Control Protocol (TCP).

Nowadays, TCP is the preferred transport protocol for YouTube and other video servers since the majority of video content delivery over the Internet is not live and most users' bandwidth is usually greater than the video coding rate. The HTTP/TCP architecture also solves the problem of access blockings carried out by many firewalls for unknown UDP ports. Additionally, the continuous improvements in latency reduction and throughput maximization achieved in new cellular technologies have allowed using TCP for minimizing the impact of errors without reducing severely the effective throughput.

The video clip download process is started by the end user when a request (with a link to the desired video clip) is sent to the YouTube web server (see Figure 1). When the client web browser receives the YouTube web page, the embedded player initiates the required signaling with the media server indicating the video to be played-out along with some setup parameters [2]. Then, the server starts progressively sending the video data over an HTTP response. The video data is then stored in a play-out buffer at the client side before being displayed. Once the download has been started, there is no further client to server signaling (unless the user interacts with the player).

The video data transfer from the media server to the client consists of two phases: initial burst of data and throttling algorithm [2]. In the initial phase, the media server sends an initial burst of data (whose size is determined by one of the setup parameters) at the maximum available bandwidth. Then, the server starts the throttling algorithm, where the data are sent at a constant rate (normally at the video clip encoding rate multiplied by the throttle factor, also denoted in the setup parameters). In a network congestion episode, the data that are not able to be delivered at this constant rate are buffered in the server and released as soon as the congestion is alleviated. When this occurs, data are sent at the maximum available bandwidth. Whenever the player's buffer runs out of data, the playback will be paused, leading to a rebuffering event.

Like quality of Internet services in general, Internet video streaming quality is mainly depending on throughput. However, quality requirements in terms of throughput are more demanding than those for other popular Internet applications as file download, web browsing and messaging. The main differences are that throughput has to meet rather precise requirements and that these requirements are stream-specific, i.e. if data are not transmitted according to playing rate (corrected by the influence of initial buffering), a rebuffering will likely occur and user QoE will drop down rapidly. It is therefore essential not only to measure the download throughput, but also to check against the bitrate the individual stream is encoded with.

There exist many quality metrics to characterize the video quality. Some of them are based on comparing the received (and degraded) video with the original video (usually called "reference"). Examples of this type of quality metrics are: Mean Square Error (MSE) [15], Peak Signal to Noise Ratio (PSNR) [15], Video Structural Similarity (VSSIM) [16], Perceptual Evaluation of Video Quality (PEVQ) [17] and Video Quality Metric (VQM) [18]. This type of metrics is useful for obtaining objective metrics in controlled experiments, but they are not applicable for online (real-time) procedures as the full reference is not available. Furthermore, they are suited to measure the image quality degradation, e.g., due to packet losses or compression algorithms. Since using TCP, packet losses are recovered, this type of metrics is less useful for YouTube.

That is why other works are oriented to provide a model for estimating the video quality without a reference. For instance, the work described in [19] presents a regression model to estimate the visual perceptual quality in terms of MOS for MPEG-4 videos over wireless networks. However, this algorithm requires an image reconstruction process to evaluate the differences between the original and the resulting images (after network transmission), which makes it not adequate for online quality estimations. In [20], the impact of delay and delay variation on user's perceived quality for video streaming is analyzed. However, it does not consider other objective metrics such as resolution, frame rate, or packet losses, which are also important for obtaining an accurate QoE estimation. In [21], a no-reference subjective metric to evaluate the video quality is presented, which considers the frame rate or the picture resolution, although their computation is complex to be used real-time.

Our implementation is based on the work presented in [22], which studied how the network QoS affects the QoE of HTTP video streaming. In this work, they propose a generic procedure to estimate the end-user's perceived quality following three steps:

1) estimate (through modeling) or measure network QoS (e.g. throughput, round trip time, loss rate, etc.);
2) convert network QoS metrics onto application QoS (application performance metrics) by means of protocols' modeling;
3) map application QoS onto end-user's QoE (in terms of MOS).

It should be noticed that the first step might not be needed if the mobile terminal is equipped with a customized YouTube's client that directly monitors and reports the application performance metrics. Otherwise, the mobile terminal shall be able to convert the network QoS onto application QoS by specific protocol modeling. For instance, there are different TCP performance models to estimate TCP throughput from network QoS [23][24]. Afterwards, application performance metrics ($T_{init}$, $f_{rebuf}$, $T_{rebuf}$) can be estimated at the receiver from performance indicators at lower layers (e.g. TCP throughput) as well as other parameters like the video coding rate, video length, buffer size at the receiver or the minimum buffer threshold that triggers a rebuffering event (see [22] for further details).

The model to estimate application QoS metrics from network QoS is valid under certain assumptions: 1) the network bandwidth, Round Trip Time (RTT) and packet loss rate are assumed to be constant during the video download; 2) the client does not interact with the video during the playback, such as pausing and forward/backward.

The third step is performed by applying a utility function for HTTP video streaming as a function of three application performance metrics:

- Initial buffering time ($T_{init}$): time elapsed until certain buffer occupancy threshold has been reached so the playback can start.
- Rebuffering frequency ($f_{rebuf}$): frequency of interruption events during the playback.
- Mean rebuffering time ($T_{rebuf}$): average duration of a rebuffering event.

The final MOS expression can be computed as [22]:

$$\text{MOS}_{\text{QoSmodel}} = 4.23 - 0.0672 \cdot L_{ti} - 0.742 \cdot L_{fr} - 0.106 \cdot L_{Tr} \quad (1)$$

being $L_{ti}$, $L_{fr}$ and $L_{tr}$ valued 1, 2 or 3 to represent the "low", "medium", and "high" levels of $T_{init}$, $f_{rebuf}$, and $T_{rebuf}$, respectively. The concrete values used to quantize previous application performance metrics can be found in [22]. From previous equation, it can be seen that the quantized rebuffering frequency ($f_{rebuf}$) metric has the highest impact on the end user's QoE, compared to the initial buffering time ($T_{init}$) and the rebuffering duration ($T_{rebuf}$). In this respect, it is reasonable to think that the perceived quality does not only depend on the pause intensity (percentage of time in the pause state), since a higher number of pauses (with lower pause durations) seems more annoying to the user.

## 3. Android application for YouTube QoE evaluation

The model for estimating YouTube QoE has been implemented as an Android application. Our QoE tool is able to run in two different modes:

1) *Intrusive mode*: the application includes an embedded video player (based on Media Player), thus having access to the content being consumed (through the YouTube API).

2) *Transparent mode*: the application runs in background, so monitoring functionalities are associated to YouTube sessions established either through the native YouTube application or through the web browser.

Our Android application includes the following modules:

- *Monitoring*: this module is responsible for monitoring network QoS parameters as well as other configuration parameters as required to estimate the application performance monitoring (listed in previous section). It makes use of the Android Networking and YouTube Data Application Programming Interface (API) to get a number of parameters associated to the session.

- *QoE estimation*: in charge of (automatically) computing the QoE of a YouTube session (in terms of MOS) from QoS parameters, according to Eq. (1).

- *QoE advices*: informs the user about possible causes that lead to a low MOS and provides some hints to improve it.

- *QoE user feedback*: allows users to qualify the session through an opinion survey. This information is used to correlate the QoE model with real user feedback.

- *QoE reporting*: this module is responsible for reporting all the performance indicators to a QoE server for post-processing purposes.

A general overview of our YouTube QoE framework is depicted in Figure 2. In addition to the MOS value automatically estimated by the application, users are requested to qualify the session (video, audio and general feedback) manually in the same MOS scale (from 1 to 5). We have used both types of QoE evaluations to validate the theoretical model proposed in [22], as well as to propose a modified function according to the results of our pilot experience.

Figure 3 shows some snapshots of our Android QoE tool in its *Intrusive mode* version, which includes the media player. Once the YouTube session is over, estimated quality results are shown to the user and reported to the QoE server.

In order to estimate the QoE, the monitoring module must collect a set of performance indicators to be subsequently mapped onto QoE. In addition to the three application performance metrics ($T_{init}$, $f_{rebuf}$, $T_{rebuf}$) required to compute the MOS, the monitoring module gathers other relevant information related to the mobile terminal, session information (date, type of player, etc.), location of the measurements or network information. All this information is reported, together with the estimated QoE and subjective quality specified by the users, to the QoE server. The complete list of parameters that are reported (from the terminal) to the QoE server are given in Table 1.

When our QoE tool runs in *Intrusive mode* (i.e. player embedded in the application), the measurement of the three application performance metrics ($T_{init}$, $f_{rebuf}$, $T_{rebuf}$) is straightforward, as the YouTube API provides access to this type of information.

However, in *Transparent mode*, the computation of these metrics is not so easy because it has to be estimated from network level metrics, as detailed in [22]. In particular, the following basic information is required: average TCP throughput, average playing rate and player buffer size. However, this type of estimation has a limitation due to the fact that, as throughput and playing rate may vary along the time, player's buffer utilization depends on the instantaneous throughput and play-out rate rather than their average values. Therefore, this approach might lead to slightly optimistic results.

Regarding the QoE advices module, its role is to analyze possible causes that provide a low QoE, and subsequently, provide particular advices to the user when certain conditions are given. As an example, Table 2 shows potential causes of low QoE, their associated evidences and advices.

## 4. YouTube QoE pilot experiment

A set of 17 users (engineers from Telefónica company) were selected to participate in a pilot experiment, which consisted in periodically testing our YouTube QoE tool (installed on different Android smartphones) during one month. Every YouTube native session were transparently monitored and evaluated in two ways: 1) automatically by the application (from QoE model previously described); 2) by the users through an online opinion survey. A total number of 1435 YouTube sessions were evaluated during the pilot. The data collected from each user device was sent to a server for post-processing purposes.

The pilot experience was carried out in Madrid (Spain), covering both rural and urban environments (as shown in Figure 4 on the left). Different colors represent the associated subjective quality (from the opinion survey) for a set of YouTube sessions. Such a survey (related to the video quality, audio quality and overall quality) was requested to be filled after each YouTube session. Figure 4 on the right show the probability density distribution of the feedback associated to video, audio and general quality.

According to the statistics collected at the QoE server, the majority of videos consumed by the users are short: near 90% of the videos shorter than 5 minutes and average duration 160 seconds (see Table 3). Regarding the video characteristics, users had free access to YouTube repository, so wide a variety of videos with different average bitrates (from 75 kbps to 3 Mbps depending on the resolution and codec) have been downloaded.

Next, statistics related to the application performance metrics (mainly referred to $T_{init}$, $f_{rebuf}$, $T_{rebuf}$, that are required to evaluate MOS), are analyzed in detail. Later, their effect on experienced quality will be described.

First, the box and whiskers plot of the parameter *Initial Buffering Time* ($T_{init}$) per technology is given in Figure 5. This non-parametric representation depicts quartiles as a box with median drawn as a vertical line inside the box. That is, 50% of values for $T_{init}$ are included in the interval inside the box. Moreover, lines extending from the boxes (*whiskers*) indicate variability outside the upper and lower quartiles. Outliers lying further 1.5 times the inter-quartile range are plotted as individual points, and those further three times that range (extreme points) are besides filled up. Average value is shown as a red cross. Box and whiskers plot can be seen as a kind of summary of the Cumulative Distribution Function (CDF) (whose estimation for $T_{init}$ is shown in Figure 6) and a graphical representation of numerical measures (some of which are presented in Table 4).

Results from Figure 5 and Figure 6 show that $T_{init}$ values for WiFi connections are lower than those for UMTS. For 3G sessions, estimated Coefficient of Variation (CV) is higher than 2, i.e., standard deviation of $T_{init}$ for UMTS connections is more than twice its average value. For WiFi, this dispersion measure is reduced to about 1.2. This comes from the fact that $T_{init}$ samples are much more concentrated around the median for WiFi sessions whereas UMTS presents higher range. The heavy tail results in a higher average located in the last quartile. In any case, 50% of the videos have experienced an initial buffering time shorter than 7 seconds. In most connections, no rebuffering is necessary, thus the median for the rebuffering frequency ($f_{rebuf}$) is 0 (see Table 5). However, in this case, $f_{rebuf}$ is higher for WiFi than for UMTS; the reason is that, although the number of pauses is smaller for WiFi (Table 6), videos were shorter (see Table 3), thus boosting the frequency of interruption events even if the mean rebuffering time ($T_{rebuf}$) is lower (Table 7).

Now, we are exploring the effect of performance indicators in the reported MOS. Figure 7 shows the initial buffering time ($T_{init}$) box and whisker plot per MOS. As shown in the results, lower $T_{init}$ values are associated to higher MOS. Although a higher feedback quality could be expected for WiFi than for UMTS, it can be observed that users do not assign a significantly lower MOS for UMTS than for WiFi (see Figure 8 and Table 8) even if, from an objective point of view, their performance is better (a summary of the three studied performance indicators can be found in Tables 9, 10 and 11). That is, although WiFi connections achieve much better QoS figures, MOS values are very similar to those of 3G. The reason for that might be that subjective users' expectations could be influenced by the type of connection being used. Hence, users might penalize the QoE of WiFi connections due to a higher expected quality.

Next, the appropriateness of the theoretical model in Eq. (1) is analyzed by evaluating the correlation between the theoretical MOS and the MOS reported by users, resulting in a correlation factor of 0.97, i.e. the coefficient of determination $R^2$ for linear regression through origin is 93.93%. Figure 9 on the left shows the difference between MOS results provided by the theoretical model (see Eq. (1)) and MOS reported from the users' opinion survey. It can be observed that the QoE model provides an estimation which falls within ±0.5 of the reported scored in 23% of sessions. In general, the model in [22] provides more pessimistic results than opinion of users as estimated MOS is lower than that reported for about 68% of sessions. A simple modification results from taking a linear regression between the modeled MOS and MOS as reported by users, yielding to:

$$\text{MOS}_{mod} = 1.1935 \cdot \text{MOS}_{QoSmodel} \qquad (2)$$

Note that this measurement indicates that MOS is about 20% higher than that given in Eq. (1). The reason could be that users could be more indulgent with wireless connections than for wired scenarios under which the original model was obtained.

Due to regression properties, the average value for the difference between MOS as obtained by (2) and that reported by users (that it, the residuals) is 0, although no symmetry around 0 exists (see Figure 9 on the right). Differences between subgroups per technology are not significant (estimated slope of 1.1995 for WiFi connections and 1.2089 for UMTS). It was explored if a multivariant regression could improve those results. Only linear regression was analyzed as a modification of numerical quantities as those proposed in [25] cannot be easily included in the multivariant procedure. The adjusted $R^2$ including all available parameters results in 90.5%, only a bit lower (90.46%) if the total rebuffering time is taken out from regression. As this value is lower than that obtained with (1), the heuristical measurement quantization proposed in [22] increased in a 20% seems to be able to predict well users expectations.

## 5. Conclusions

This work has presented a QoE evaluation tool for Android terminals that is able to estimate the QoE (in terms of MOS) for YouTube service based on theoretical models. In particular, this tool makes it possible to map network QoS onto the QoE of YouTube sessions. Additionally, a QoE advices module analyzes possible causes that lead to low QoE, and subsequently, provide particular advices to the user under certain conditions.

Our application has been tested on a pilot experience over 17 Android terminals during one month. According to the statistics, most of the responses from the users' survey match up with theoretical estimations; however, the QoE model provides slightly more pessimistic results than the opinion of the wireless users, probably as the model was initially generated under wired scenarios. In that sense, we propose a modified utility function from taking a linear regression between the theoretical MOS and the MOS reported by users.

In our opinion, it is critical that application developers provide access to the main Key Performance Indicators (KPIs) associated to their services in order to ease the evaluation and analysis of the QoE.

**Acknowledgements**

This work has been partially supported by the Spanish Government (TEC2010-18451).

**Figure captions**

Figure 1. Signaling flowchart of a YouTube session (via web browser).

Figure 2. YouTube QoE framework.

Figure 3. Snapshots of our Android QoE tool (Intrusive mode).

Figure 4. Geographical distribution of the users (left) and quality feedback distribution (right).

Figure 5. Box and whiskers plots for the initial buffering time (in seconds) per technology.

Figure 6. Estimated CDF of the initial buffering time ($T_{init}$).

Figure 7. Box and whisker plot for $T_{init}$ per reported MOS.

Figure 8. Box and whisker plot for MOS per used technology.

Figure 9. Histogram for the difference between the original QoS model (left) or modified QoS model (right) and as reported by users

# Tables

Table 1. List of parameters that are reported (from the terminal) to the QoE server.

| Parameter Type | Parameter | Description |
| --- | --- | --- |
| Device ID | IMEI | International Mobile Equipment Identity (15 digit format). |
| Session information | ReproductionMode | It indicates the application used for video reproduction:<br>1 - Embedded video player (based on Media Player)<br>2 - YouTube native application<br>3 - Web Browser |
| | ReproductionTime | Total reproduction time (in ms) including rebuffering and user originated pauses. |
| | Date | Date of the YouTube video session (AAAA-MM-DD). |
| | Hour | Hour of the YouTube video session (HH:MM:SS). |
| Application Performance Metrics | InitialBufferingTime ($T_{init}$) | Total time (in ms) since the user starts the session until the video is ready to be played. |
| | RebufferingFrequency ($f_{rebuf}$) | Frequency of interruption events (not forced by the user) during the playback. |
| | MeanRebufferingTime ($T_{rebuf}$) | Average duration of a rebuffering event (in ms). |
| Location of the measurement | Latitude | Expressed in sexagesimal degrees (-90, 90). |
| | Longitude | Expressed in sexagesimal degrees (-180, 180). |
| | Altitude | Expressed in meters above sea level. |
| | Accuracy | Precision in meters of the location measurements. |
| | Time | Moment at which the location measurement was done (AAAA-MM-DD_HH:MM:SS format). |
| | Provider | Method to perform location measurements: GPS or Network-assisted. |
| Network information | ConnectionType | Type of network data connection active for the session. Possible values: 0(WIFI), 1(GPRS), 2(EDGE), 3(UMTS), 4(CMDA), 5(EVDO_0), 6(EVDO_A), 7(1XRTT), 8(HSDPA), 9(HSUPA), 10(HSPA), 11(IDEN), 12(EVDO_B), 13(LTE), 14(EHRPD), 15(HSPAP) |
| | LAC | Location Area Code where the user is located. |
| | CellID | Identifier of the cell providing service to the terminal. |
| | RSSI | Received Signal Strength Indication (dBm) measured by the terminal (for either WiFi or cellular connections). |
| Subjective Quality (feedback from users) | VideoQualityFeedback | Subjective opinion regarding video quality (scale: 1 to 5). |
| | AudioQualityFeedback | Subjective opinion regarding audio quality (scale: 1 to 5). |
| | GeneralFeedback | General feedback from the user (scale: 1 to 5). |
| | AdditionalComments | The user can add any additional comment. |
| Subjective Quality (estimated) | EstimatedVideoQuality | Estimated video quality from QoE model (scale: 1 to 5) |

Table 2. Examples of causes of low QoE and advices to users (QoE advices module).

| Cause | Evidence | Advice |
|---|---|---|
| Low throughput | *High traffic load* <br>  IF many applications synchronizing → <br>  ELSE IF many apps running → <br>  ELSE → <br><br> *Low Network Traffic and connected to a cellular network* <br>  IF GSM / 3G Lock on 2G → <br>  ELSE IF low RSSI & WiFi available → <br>  ELSE IF low RSSI & WiFiSwitchedOn <br>    & WiFi Not Available → <br>  ELSE IF low RSSI & Bluetooth <br>    Switched On → <br><br> *Low Network Traffic and connected to WiFi* <br>  IF WiFi Tethering is activated → <br>  ELSE IF Bluetooth Switched On → <br>  ELSE → | Temporarily stop data synchronization <br> Offer some apps/services to be switched off <br> Switch to other technology (WiFi, Mobile) <br><br><br> Activate 3G <br> Switch to a WiFi connection <br><br> Switch off WiFi to avoid interference <br><br> Switch off Bluetooth to avoid interference <br><br><br> Switch off WiFi Tethering <br> Switch off Bluetooth <br> Switch to a cellular network connection |
| Low memory | *Low Memory status flag is TRUE* <br>  IF many apps/services running → <br>  ELSE IF "hungry" app detected → <br>  ELSE → | Offer some apps/services to be switched off <br> Offer to switch off "hungry" app <br> Check for system updates |
| High CPU load | CPU load is high during a period <br>  IF many apps/services running → | Offer some apps/services to be switched off |
| Low CPU frequency forced | *CPU freq low* <br>  IF low battery level <br>    OR high battery temperature → <br>  ELSE IF aggressive power save <br>    profile selected → <br>  ELSE → | Wait until battery gets in better conditions <br><br> Select a performance oriented profile <br> Check for system updates |
| Video requirements exceeds terminal capabilities | *YouTube API video source and device HW information* <br>  IF device capability < video req. → | Try to select less demanding video files, switch off High Quality (HQ) option. |
| Low video quality in origin | *YouTube API video source information* <br>  IF low resolution/codingRate → | Select another file of higher quality |

Table 3. Summary for Reproduction Time (in minutes).

| Connection Type | No. sessions | Mean | Median | Standard Deviation | Min | Max |
|---|---|---|---|---|---|---|
| UMTS | 911 | 2.91 | 1.8 | 3.92 | 0.02 | 49.35 |
| WiFi | 524 | 2.19 | 1.14 | 2.89 | 0.0007 | 26.88 |
| Total | 1435 | 2.65 | 1,6 | 3.6 | 0.0007 | 49.35 |

Table 4. Summary for initial buffering time time (Tinit) in seconds.

| Connection Type | Mean | Median | Standard Deviation | Min | Max |
|---|---|---|---|---|---|
| UMTS | 15.24 | 7.604 | 31.8225 | 2.69 | 582.733 |
| WiFi | 7.94 | 6.014 | 9.69786 | 2.389 | 154.245 |
| Total | 12.5758 | 6.977 | 26.2543 | 2.389 | 582.733 |

Table 5. Summary for rebuffering frequency (frebuf)

| Connection Type | Mean | Median | Standard Deviation | Min | Max |
|---|---|---|---|---|---|
| UMTS | 1.88e-3 | 0 | 5.55e-3 | 0 | 46e-3 |
| WiFi | 1.05e-3 | 0 | 6.73e-3 | 0 | 0.11 |
| Total | 1.57e-3 | 0 | 6e-3 | 0 | 0.11 |

Table 6. Summary for number of pauses (Npauses)

| Connection Type | Mean | Median | Standard Deviation | Min | Max |
|---|---|---|---|---|---|
| UMTS | 0.54 | 0 | 2.43 | 0 | 31 |
| WiFi | 0.21 | 0 | 1.71 | 0 | 36 |
| Total | 0.42 | 0 | 2.20 | 0 | 36 |

Table 7. Summary for Mean Rebuffering Time (Trebuf) in seconds

| Connection Type | Mean | Median | Standard Deviation | Min | Max |
|---|---|---|---|---|---|
| UMTS | 19.1 | 0 | 143.2 | 0 | 2594 |
| WiFi | 2,46 | 0 | 28.75 | 0 | 632 |
| Total | 13,04 | 0 | 115.73 | 0 | 2594 |

Table 8. Percentage of reported MOS per technology

| Connection Type | MOS=1 | MOS=2 | MOS=3 | MOS=4 | MOS=5 | Average | Standard Deviation |
|---|---|---|---|---|---|---|---|
| UMTS | 3.86 | 8.13 | 26.06 | 34.55 | 27.44 | 3.74 | 1.1 |
| WiFi | 3.4 | 7.46 | 24.41 | 46.78 | 21.02 | 3.81 | 0.87 |
| Total | 2.69 | 8.18 | 25.64 | 38.46 | 25.03 | 3.74 | 1 |

Table 9. Average, max and standard deviation values (in seconds) for Tinit as per technology and MOS.

| $T_{ini}$ (s) | UMTS | | | | WiFi | | | | Average | Standard Deviation |
|---|---|---|---|---|---|---|---|---|---|---|
| | Ave | Min | Max | Std | Ave | Min | Max | Std | | |
| MOS=1 | 98,44 | 6,05 | 439,2 | 99,91 | - | - | - | - | 87,94 | 96,53 |
| MOS=2 | 22,9 | 3,15 | 147 | 29,86 | 13,54 | 2,87 | 106 | 21,59 | 20,36 | 27,55 |
| MOS=3 | 16,77 | 3,21 | 167 | 24,84 | 8,27 | 2,39 | 60,68 | 8,04 | 14,67 | 21,83 |
| MOS=4 | 11,31 | 2,69 | 83,2 | 11,21 | 7,67 | 2,66 | 154,2 | 12,99 | 9,87 | 12,39 |
| MOS=5 | 9,30 | 3,33 | 94,7 | 9,85 | 7,00 | 2,49 | 35,50 | 5,11 | 8,61 | 8,57 |

Table 10. Average, max and standard deviation values for frebuf as per technology and MOS (all min values are 0).

| $f_{rebuf}$ | UMTS | | | WiFi | | | Average | Standard Deviation |
|---|---|---|---|---|---|---|---|---|
| | Ave | Max | Std | Ave | Max | Std | | |
| MOS=1 | 13e-3 | 0,03 | 8.7e-3 | - | - | - | 0.013 | 9e-3 |
| MOS=2 | 3.6e-3 | 0.025 | 6.8e-3 | 1.4e-3 | 0.016 | 4.13e-3 | 2.8e-3 | 6e-3 |
| MOS=3 | 4.3e-3 | 0.046 | 8.6e-3 | 2.1e-3 | 0.1 | 13e-3 | 3.5e-3 | 10.4e-3 |
| MOS=4 | 1.2e-3 | 0.022 | 3.8e-3 | 6e-5 | 5.5e-3 | 5.4e-4 | 6.6e-4 | 2.9e-3 |
| MOS=5 | 0 | 0 | 0 | 7e-5 | 4e-3 | 5.4e-4 | 2e-5 | 3.1e-4 |

Table 11. Average, max and standard deviation values (in seconds) for mean rebuffering time as per technology and MOS (all min values are 0).

| $T_{rebuf}$ | UMTS | | | WiFi | | | Average | Standard Deviation |
|---|---|---|---|---|---|---|---|---|
| | *Ave* | *Max* | *Std* | *Ave* | *Max* | *Std* | | |
| MOS=1 | 520.88 | 2593.56 | 768.38 | - | - | - | 494,84 | 756,9 |
| MOS=2 | 28.26 | 372.42 | 71.18 | 4,28 | 53,03 | 12,73 | 19,75 | 58,56 |
| MOS=3 | 11.25 | 164.127 | 32.97 | 0,99 | 59,21 | 7,02 | 7,56 | 27,13 |
| MOS=4 | 2.71 | 98.056 | 11.68 | 0,11 | 9,07 | 0,92 | 1,54 | 8,78 |
| MOS=5 | 0 | 0 | 0 | 0,23 | 14,07 | 1,79 | 0,07 | 1,00 |

**Biographies**

Gerardo Gómez received his B.Sc. and Ph.D degrees in Telecommunications Engineering from the University of Málaga (Spain) in 1999 and 2009, respectively. From 2000 to 2005 he worked at Nokia Networks and Optimi Corporation (recently acquired by Ericsson), leading the area of QoS for 2G and 3G cellular networks. Since 2005, he is an associate professor at the University of Málaga. His research interests include the field of mobile communications, especially QoS/QoE evaluation for multimedia services and radio resource management strategies for LTE and LTE-Advanced.

Lorenzo Hortigüela is a Telecommunication engineer from E.T.S.I.T – UPM, specialized in computer science. He has worked mainly in defense aerospace industry as military avionics engineer and in wireless telecommunication industry as systems engineer. And he also worked for public administration as a consultant in computer science and also as a contract reviewer. Since 1996 he has worked for Telefónica I+D, always involved in mobile network R&D. Recently, he worked at the area of PDI - Enabling Platforms (SLA) as a computer technology specialist. Currently, he is working at the area of PTI - Capacity & Traffic Analysis & Solutions, developing tools based on knowledge extraction, machine learning and analytic prediction techniques.

Quiliano Perez is Physicist for the Complutense University of Madrid from 1986. In 1987 he started working in Telefónica I+D, in aspects related with reliability predictions, failure analysis and components electronic testing, used in telecommunication systems. In 2011 he was working in QoE issues for mobile broadband networks, analyzing and developing tools and client applications to control and manage the QoS and QoE of mobile services.

Currently, he is working in GCTO (Global Chief Technology Office) of Telefónica, analyzing and defining the requirements needed for residential cellular gateways.

Javier Lorca received his B.Sc. in Telecommunications in 1998 by the Universidad Politécnica de Madrid. In 1999 he worked for Teldat on ciphering techniques, and since 2000 he is working in Telefónica I+D on several areas related with mobile communications and physical layer performance, involving terminals specifications and testing, link-level and system-level simulations of 3G/3.5G systems, LTE digital signal processing, and LTE-Advanced. He is currently working on Quality of Experience in wireless networks.

Raquel García received her B.Sc. Degree in Telecommunication Engineering in 1998 from Madrid Polytechnics University. Between 1997 and 1998 she staged at the Digital Communication Systems Department at the Technical University Hamburg-Harburg (Germany), where she carried out her master thesis about multipath fading channel modeling. In July 1998 she signed for Telefónica I+D, becoming part of the Radio Communication Systems Department. Her carrier has oriented towards mobile communications, especially on radio planning and optimization and QoE in LTE.

Mari Carmen Aguayo-Torres received the M.S. and Ph.D. degrees in Telecommunication Engineering from the University of Malaga, Spain, in 1994 and 2001, respectively. Currently, she is working at the Department of Communications Engineering, at the same university. Her main research interests include adaptive modulation and coding for fading channels, multi-user OFDM, SC-FDMA, cross-layer design and probabilistic QoS guarantees for wireless communications.

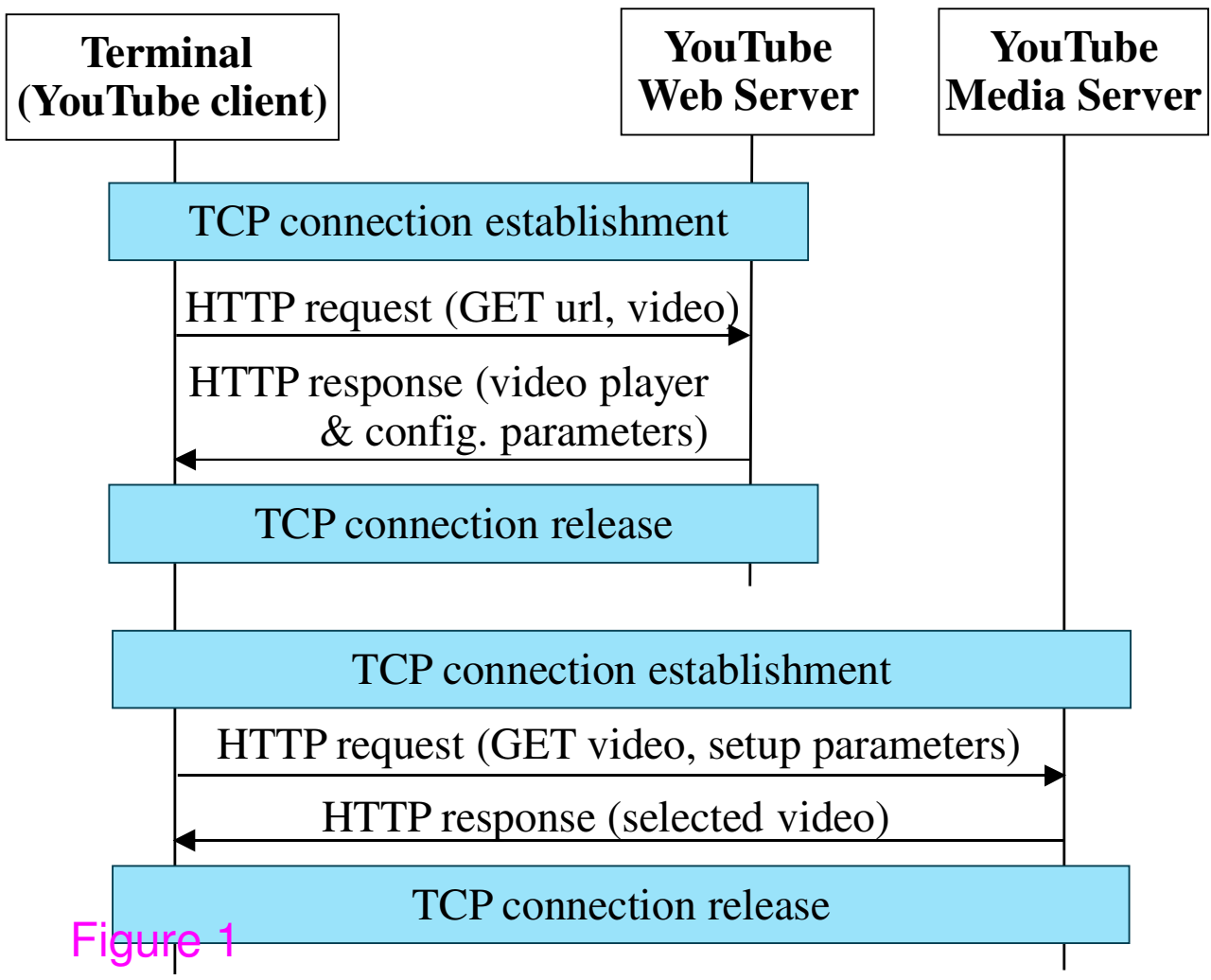

Figure 1

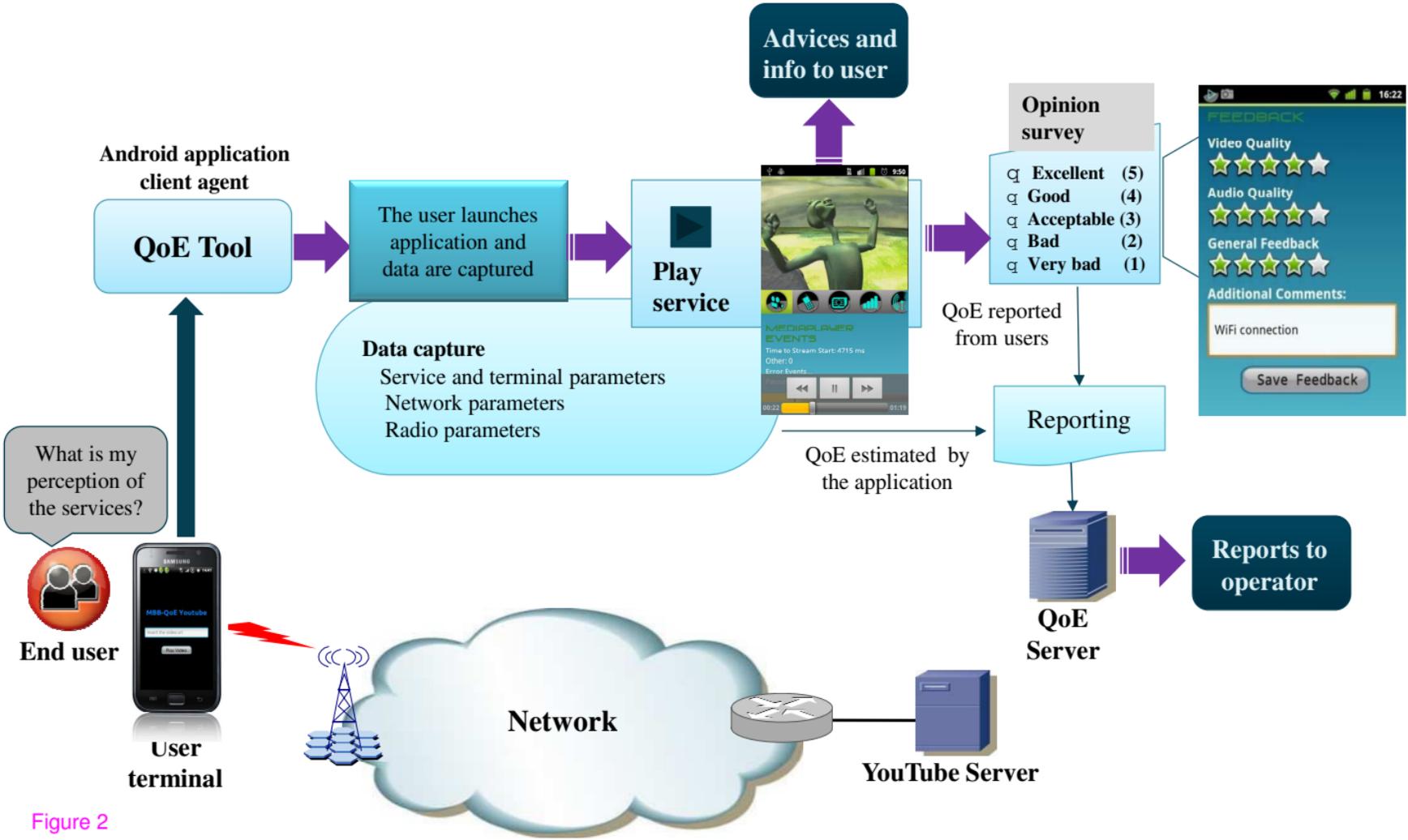
Figure 2

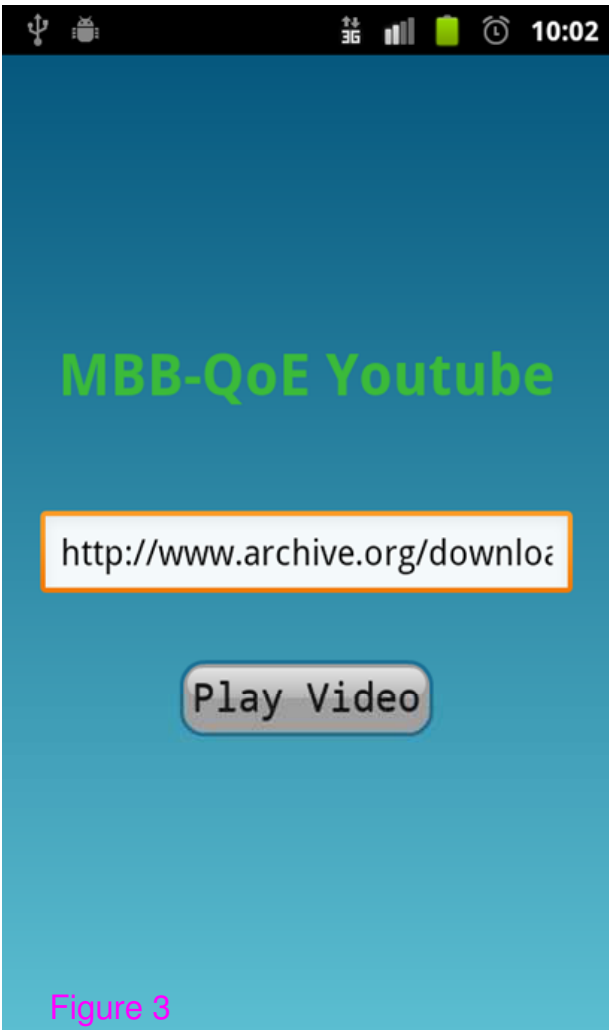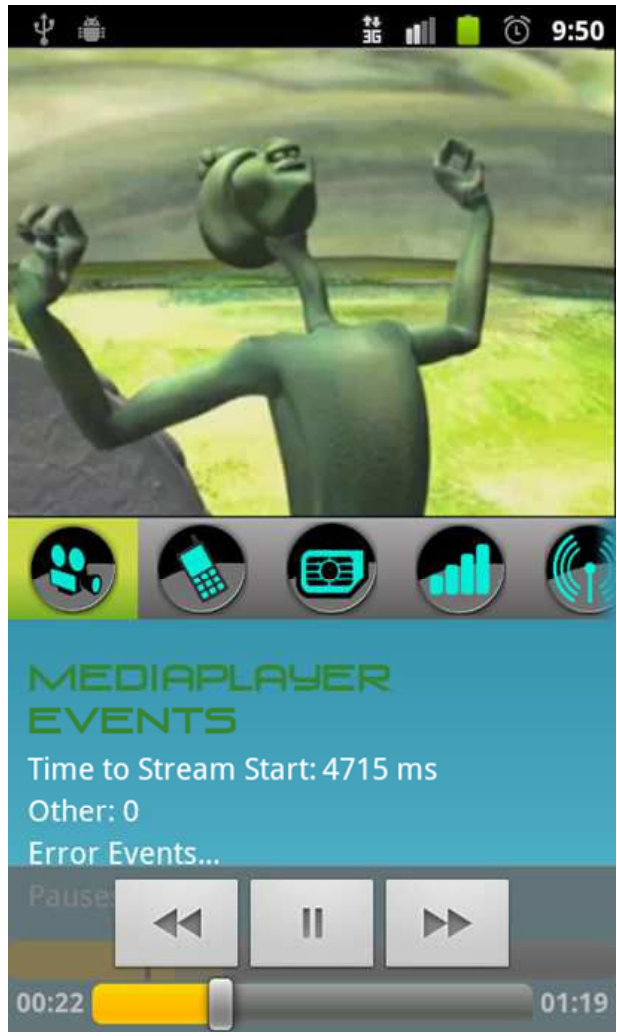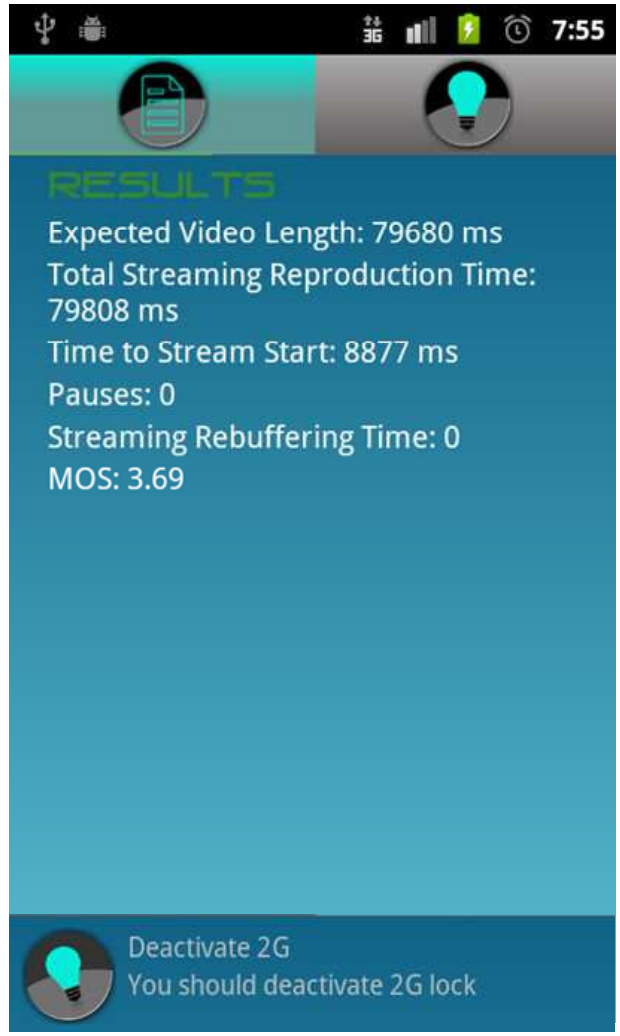

Figure 3

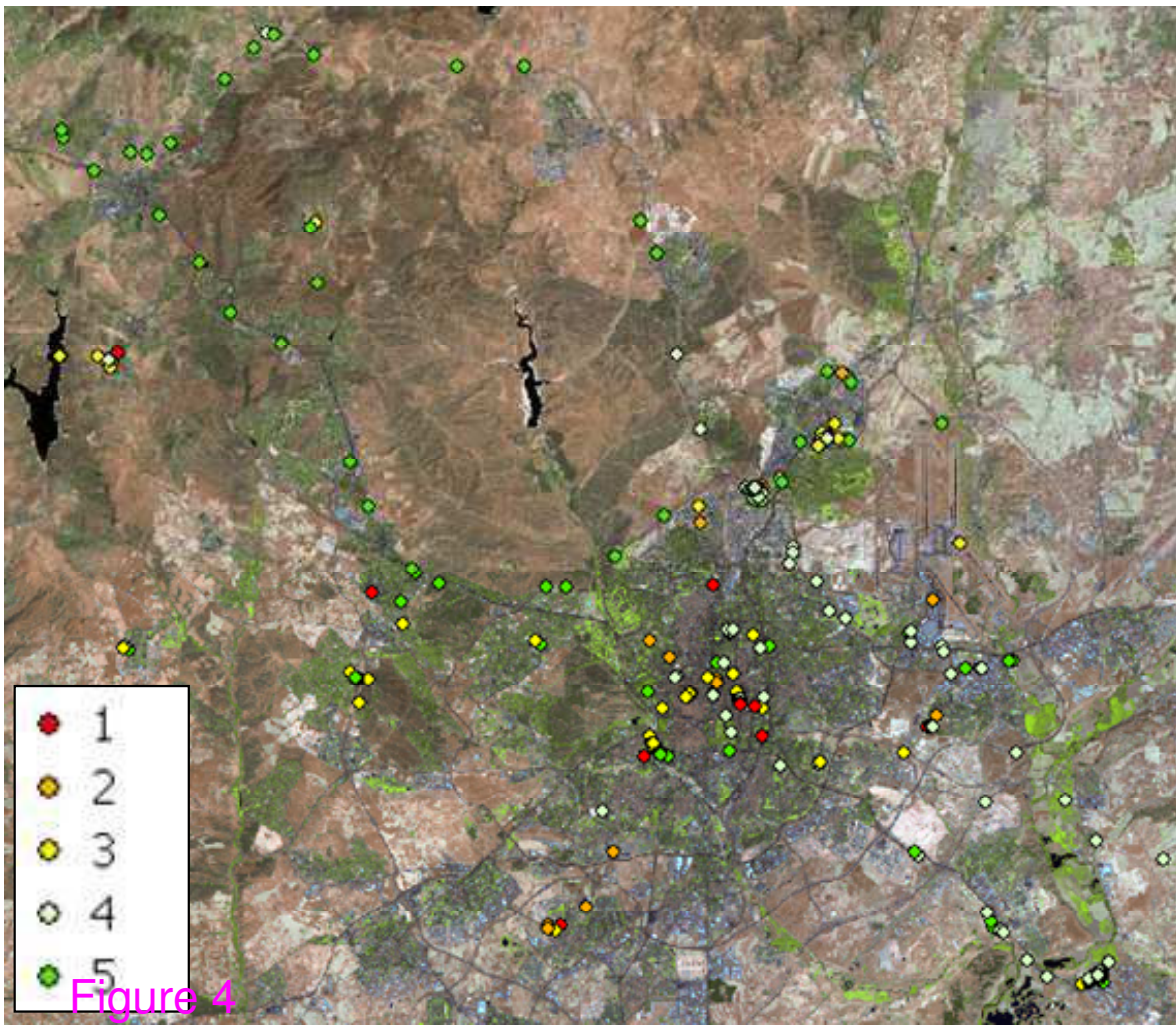
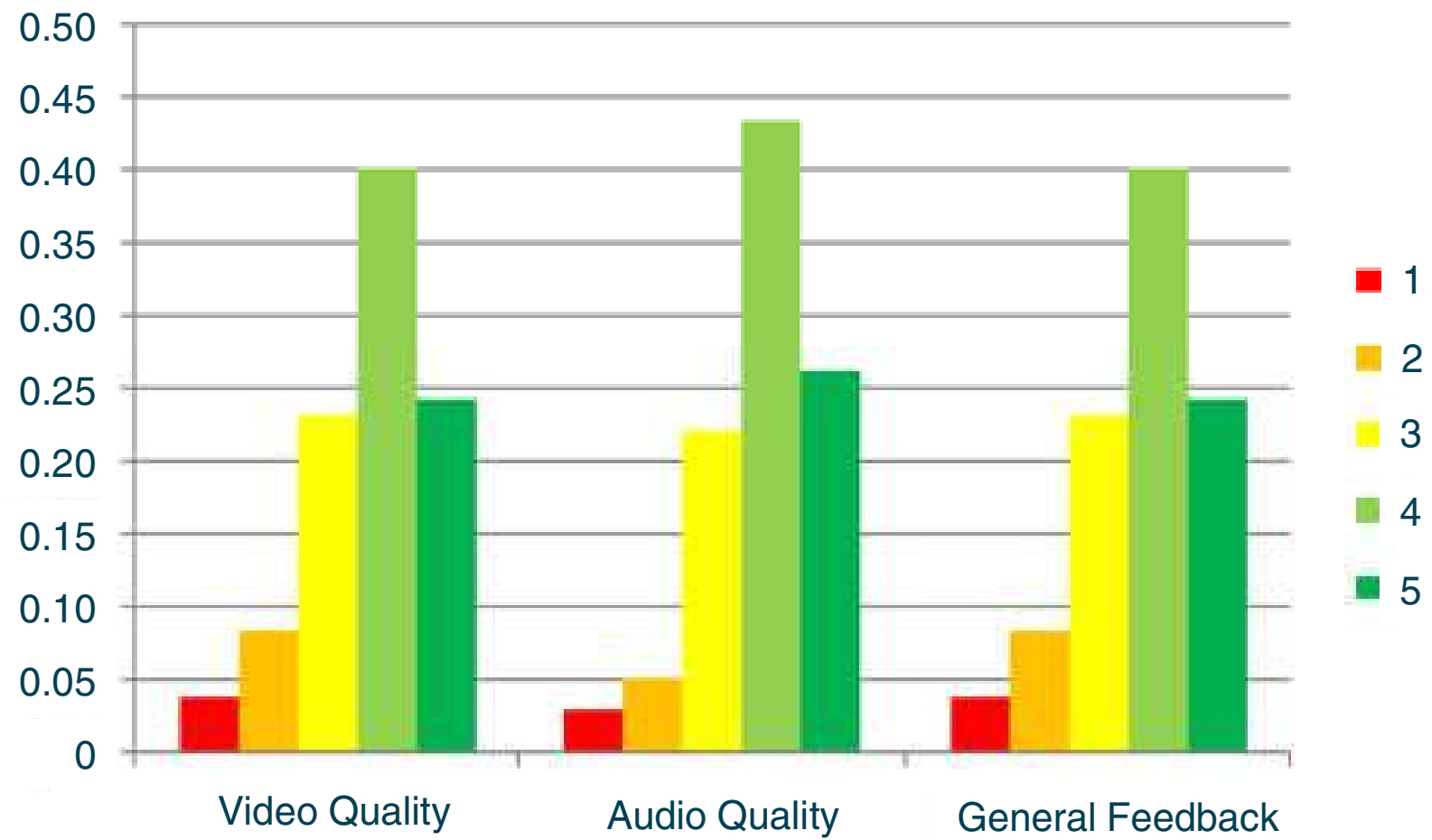

Figure 4

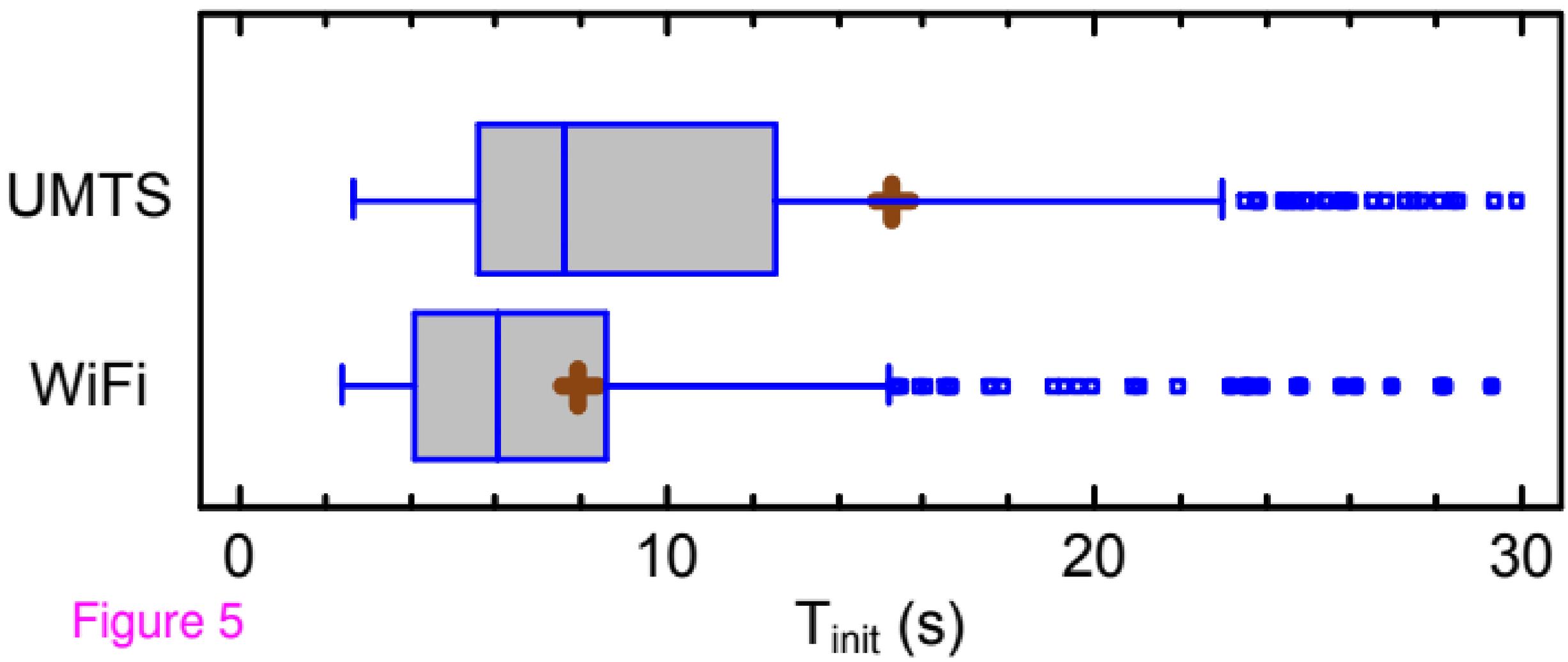

Figure 5

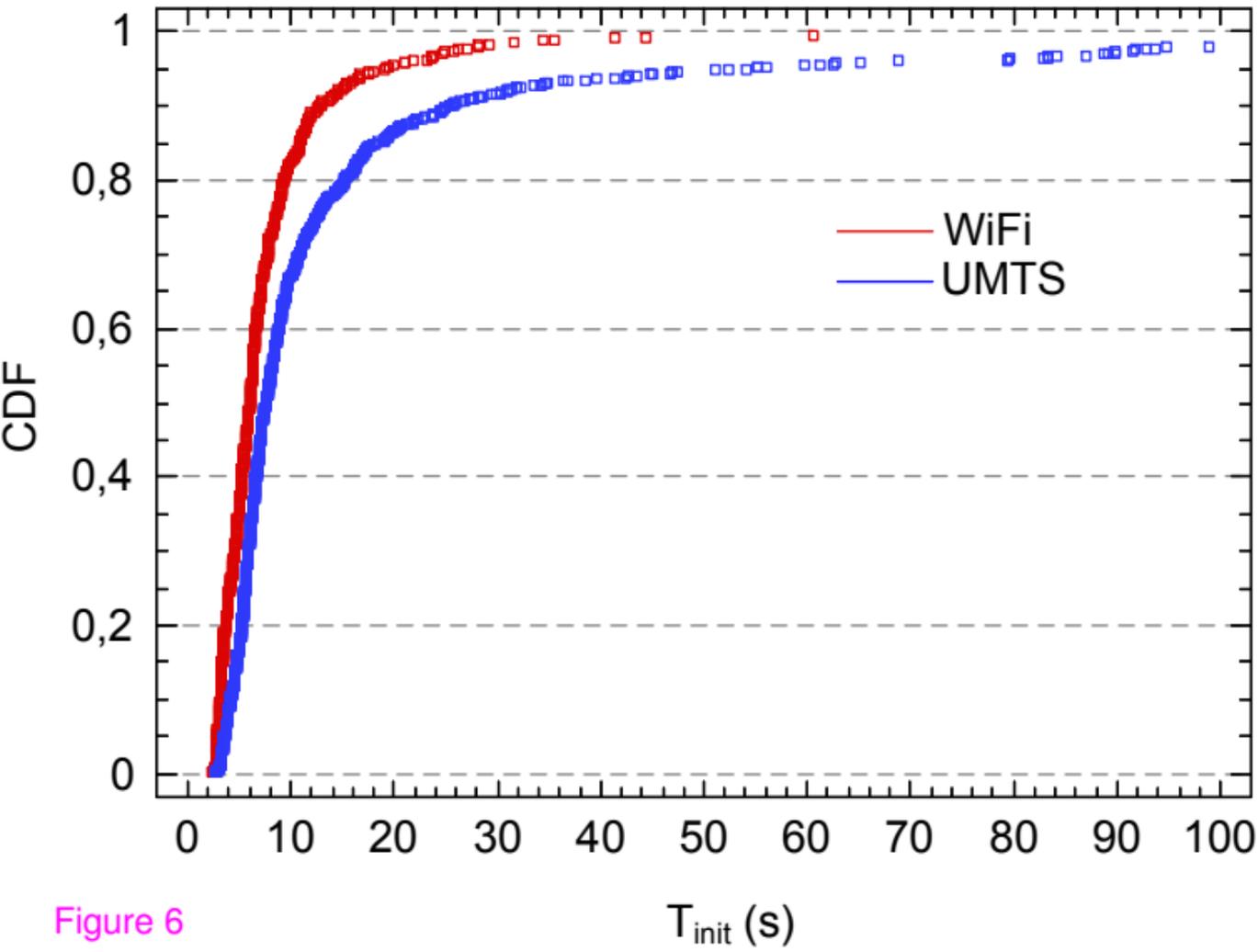

Figure 6

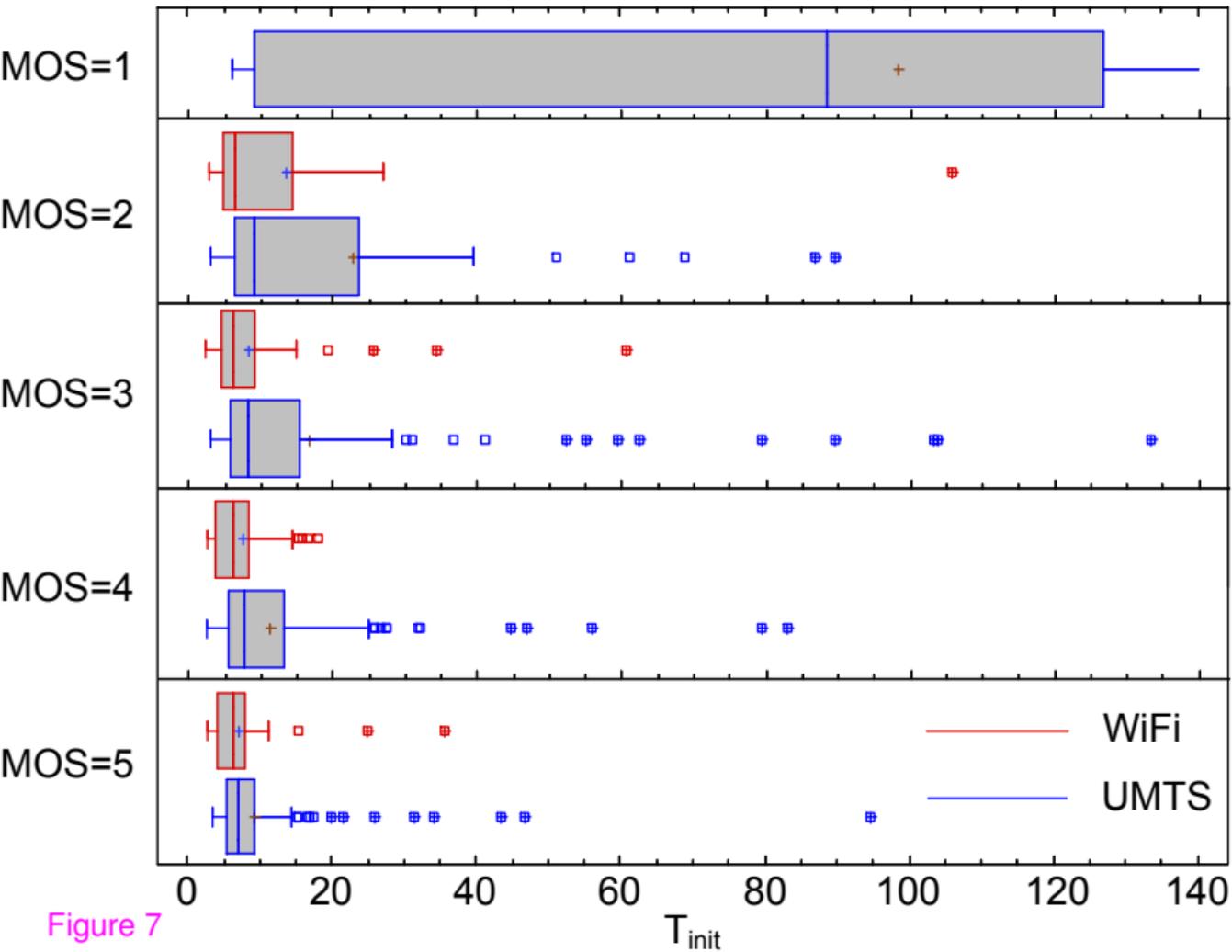

Figure 7

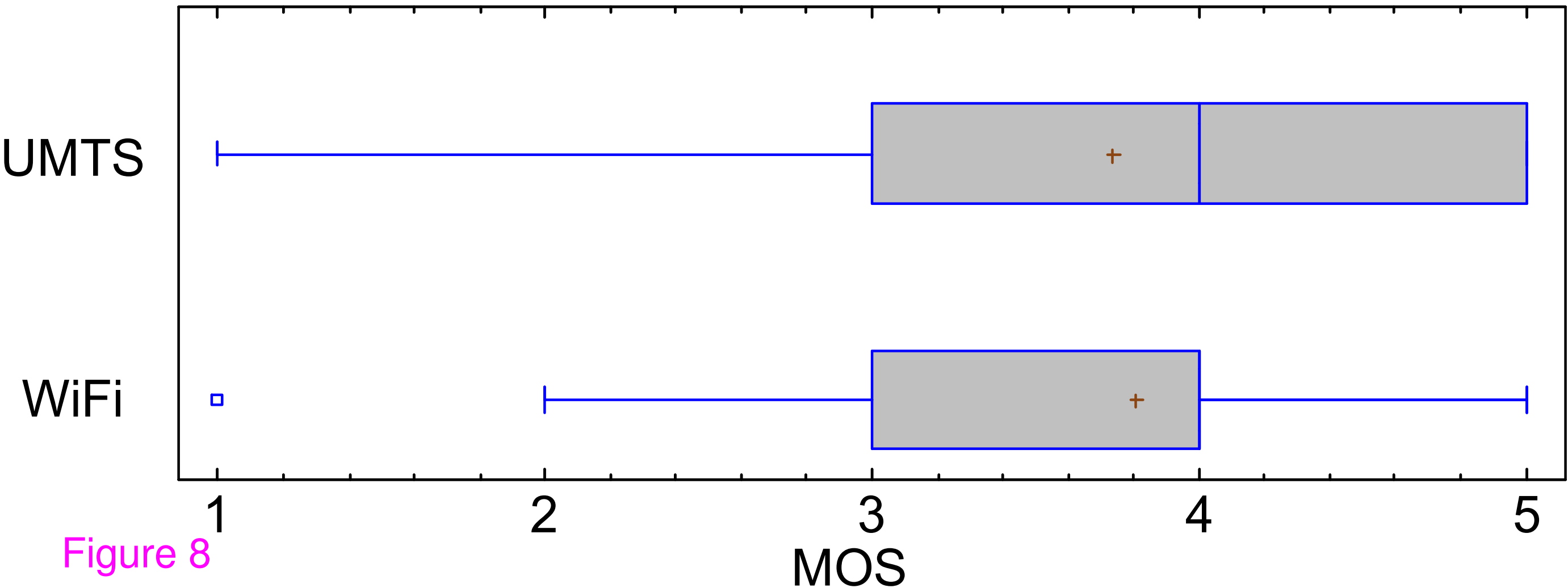

Figure 8

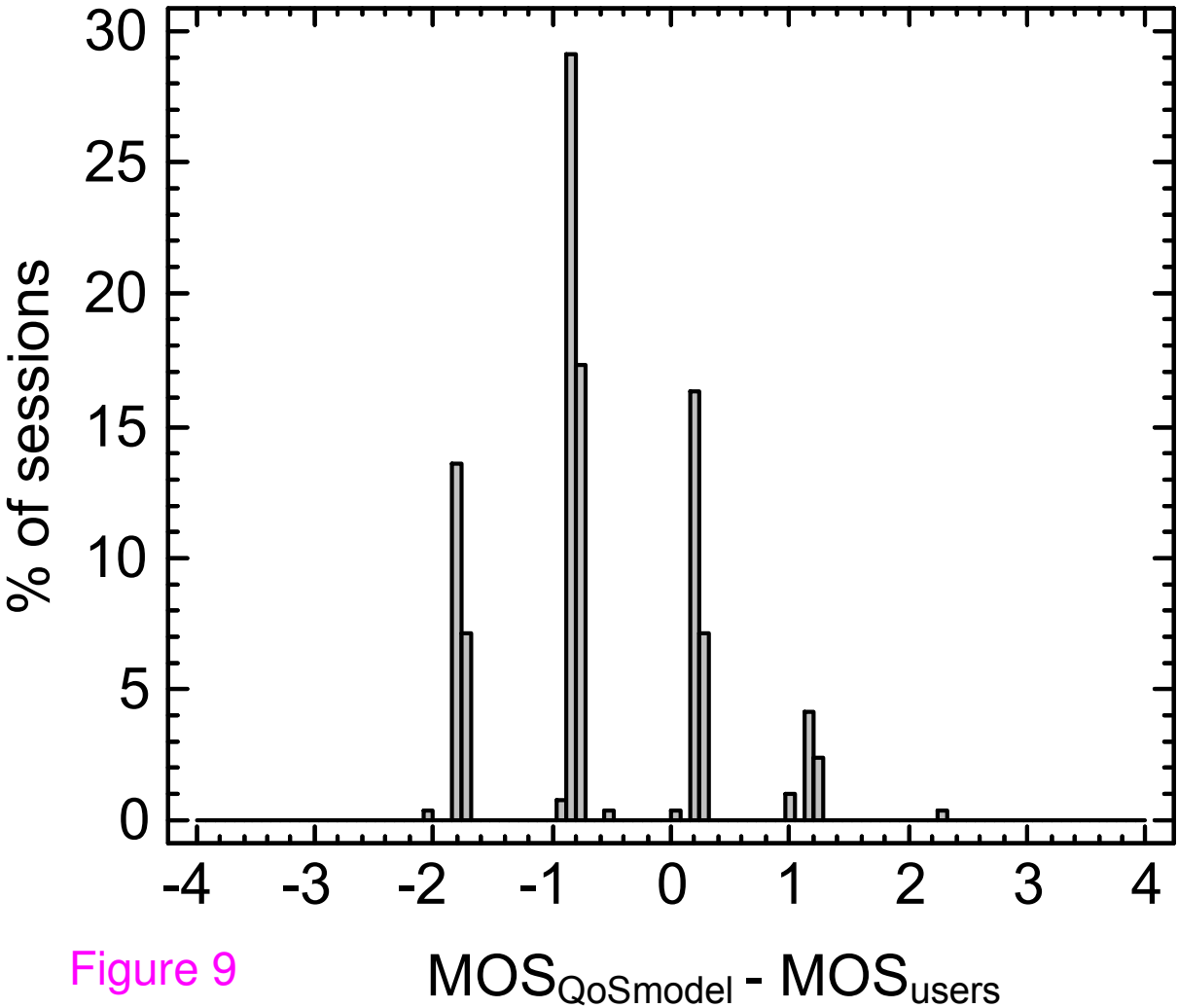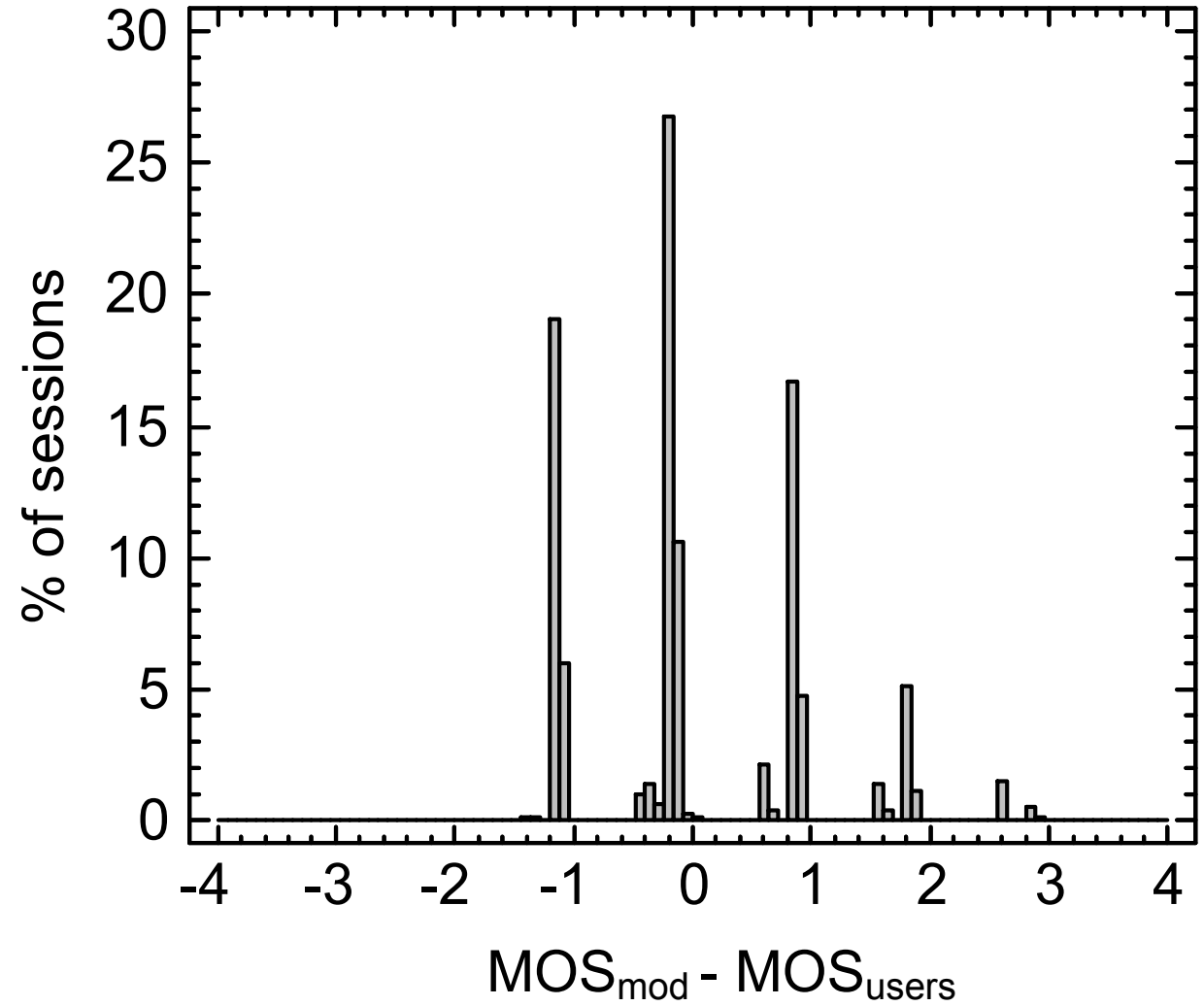

Figure 9